\documentclass[prd,twocolumn,nofootinbib]{revtex4}
\usepackage{amsmath}
\usepackage{amssymb}
\usepackage{graphicx}
\usepackage[usenames]{color}
\usepackage{mathrsfs} 
\newcommand{\scri}{{\mathscr I}}

\begin{document}

\title{Scalar Fields in Black Hole Spacetimes}

\author{Izak Thuestad} 
\affiliation{Department of Physics, University
  of Massachusetts, Dartmouth, MA 02747}

\author{Gaurav Khanna} 
\affiliation{Department of Physics, University
  of Massachusetts, Dartmouth, MA 02747}

\author{Richard H.~Price} \affiliation{Department of Physics, MIT, 77 Massachusetts Ave., Cambridge, MA 02139}
\affiliation{Department of Physics, University  of Massachusetts, Dartmouth, MA 02747}

\begin{abstract}
The time-evolution of matter fields in black hole exterior spacetimes is a well-studied subject, 
spanning several decades of research. However, the behavior of fields in the black hole interior 
spacetime, has only relatively recently begun receiving some attention from the research 
community. In this paper, we numerically study the late-time evolution of scalar fields 
in both Schwarzschild and Kerr spacetimes, including the black hole interior. We recover the 
expected late-time power-law ``tails'' on the exterior (null infinity, time-like infinity 
and the horizon). In the interior region, we find an interesting oscillatory behavior that is 
characterized by the multipole index $\ell$ of the scalar field. In addition, we also study the 
extremal Kerr case and find strong indications of an instability developing at the horizon.    
\end{abstract}

\maketitle

\section{Introduction}\label{sec:overview}

Study of fields in black hole (exterior) spacetimes has been a popular area of research over 
many decades. Such research has often yielded many intriguing aspects of black hole physics. 
For instance, the notion of quasi-normal ringing as a characteristic feature of such spacetimes 
recently helped LIGO discover gravitational waves from a binary black hole system~\cite{GWPRL1,GWPRL2} 
for the first time ever.  

Another interesting feature of physical fields evolving in black holes spacetimes is their 
late-time, power-law decay behavior -- so-called ``tails''. This discovery was made by 
Richard Price nearly half-a-century ago in the context of a Schwarzschild black hole~\cite{price}. 
However, it has continued to be an area of active research in the Kerr black hole context~\cite{barack,hod} 
even as recently as the last few years~\cite{tails}.   

Naturally, nearly all the focus of research in this area happens to be in the context of the 
black hole exterior spacetime. However, recently interest in the interior spacetime has also 
been increasing and a number of intriguing results have emerged~\cite{ori,brady,cauchy,eilon,others}. 

In this paper, we study the late-time behavior of scalar fields (to linear order) over the entire 
black hole spacetime, including the interior region. We consider both Schwarzschild and Kerr black 
holes for this study. Late-time tail results for the exterior black hole spacetimes are known, and 
we compare our results with those (horizon, timelike and null infinity) for validation purposes. 
The main new results we present in this work relate to the late-time behavior of scalar fields
in the black hole interior region. In order to perform such a study, we make use of advanced  
mathematical (hyperboloidal compactification) and computational techniques (high-precision
GPU-computing), the details of which appear in the following sections. In this work, we focus 
on axisymmetric, scalar field configurations. Other, more general cases will be presented elsewhere. 

Our work involves using compactified ingoing Kerr coordinates to perform computations that cover 
the entire interior and exterior black hole spacetime. These coordinates were successfully used 
in recent work to perform a detailed study of the Cauchy horizon~\cite{cauchy} of a rotating black hole. 
We find that while there is a way to view the late-time decay of the fields in the black hole interior 
as a power-law tail, an infalling physical detector would actually record a finite number of oscillatory 
cycles\footnote{These are different from the oscillations on the Cauchy horizon as found by Ori in 
1992~\cite{ori92} in the context of a Kerr black hole.} before encountering the spacetime singularity. 
The number of oscillatory cycles depends on the multipole index of the field, and also whether the 
black hole is spinning or not.  

We also include a short section on the late-time behavior of scalar fields at the horizon of 
an extremal Kerr black hole, and report on an indication of the formation of an asymptotic  
instability, as very recently uncovered in the research literature~\cite{aretakis,gralla}. 

This paper is organized as follows: Section II offers details on our methodology, i.e. 
the approach we take in solving the scalar Teukolsky equation using suitable coordinates 
that allow us to ``penetrate'' the black hole horizon and continue to evolve the fields into 
the interior region; Section III documents our numerical results and the comparisons with 
expectations based on previous work; and finally, we end with a brief summary and 
statement on future work in Section IV.

\section{Numerical Solution of the Teukolsky Equation}

In this section we briefly document the background and computational methods 
used to generate the results in the upcoming sections of this paper. We provide 
here a description of the coordinate-systems used, the relevant evolution equations 
and the computational techniques employed.   

The main context of this work is the behavior of scalar fields in the spacetime 
of a rotating black hole (Kerr), including the special case of a non-rotating 
(Schwarzschild) hole. The common coordinate system that is used to describe 
the spacetime of black holes is the Boyer-Lindquist system $(t,r,\theta,\varphi)$ 
that has close similarities with spherical coordinates. However, Boyer-Lindquist 
coordinates are not the best suited to study the black hole interior spacetime 
because they suffer from a coordinate-singularity at the horizon locations. Since 
we are interested in studying the behavior of fields in the interior region too, we 
instead make use of a better suited coordinate system, i.e. ingoing Kerr coordinates. 
These are a Kerr spacetime generalization of the better-known Eddington-Finkelstein 
coordinates that are able to smoothly ``penetrate'' the horizon of a Schwarzschild 
black hole. In the following subsection, we review the relationship between these 
different coordinate systems and emphasize some of their important aspects.  

The main evolution equation of interest in this work is the Teukolsky master 
equation that describes scalar, vector and tensor field perturbations in the 
spacetime of a Kerr black hole~\cite{teuk} to linear order. We numerically solve this 
equation for the scalar field case using a compactified form of the ingoing Kerr coordinates. 
Using  hyperboloidal compactification allows us to {\em directly} sample the behavior of 
fields throughout the spacetime, including even null infinity $\scri^+$. One important 
aspect of this work worth noting is that we must evolve the fields for a long duration 
because we are interested in the late-time, power-law decay behavior of the fields. 
This behavior typically appears {\em after} the quasi-normal modes of the system have 
exponentially decayed enough to become subdominant. This posed certain challenges that 
are explained in some detail in the following sections.  

The following subsections offer additional details including the main expressions 
for the quantities involved and also our computational methodology. 

\subsection{Teukolsky Equation in Ingoing-Kerr Coordinates}

We begin with an expression of the usual Boyer-Lindquist coordinate version of 
the Kerr spacetime metric and the associated Teukolsky equation~\cite{teuk}. The 
metric has the form  
\begin{eqnarray}
&&
ds^2=\left(1-2Mr/\Sigma\right) dt^2
\nonumber\\
&&
+\left(4Mar\sin^2\theta/\Sigma\right)dtd\varphi 
-\left(\Sigma\over \Delta\right) dr^2 - \Sigma d\theta^2
\nonumber\\
&&
-\sin^2\theta\left(r^2+a^2+2Ma^2r\sin^2\theta/\Sigma\right) d\varphi^2,
\end{eqnarray}
where $\Sigma=r^2+a^2\cos^2\theta$ and $\Delta=r^2-2Mr+a^2$. Here $M$ refers 
to the black hole mass and the Kerr parameter $a$ characterizes the hole's spin. 
It is clear that the metric exhibits pathological behavior at the horizon 
locations, i.e. when $\Delta=0$. Note that this coordinate singularity can be 
easily removed by a suitable change of coordinates. The Teukolsky master equation 
takes the form
\begin{eqnarray}
&&
-\left[\frac{(r^2 + a^2)^2 }{\Delta}-a^2\sin^2\theta\right]
         \partial_{tt}\Psi
-\frac{4 M a r}{\Delta}
         \partial_{t\phi}\Psi \nonumber \\
&&- 2s\left[r-\frac{M(r^2-a^2)}{\Delta}+ia\cos\theta\right]
         \partial_t\Psi\nonumber\\  
&&
+\,\Delta^{-s}\partial_r\left(\Delta^{s+1}\partial_r\Psi\right)
+\frac{1}{\sin\theta}\partial_\theta
\left(\sin\theta\partial_\theta\Psi\right)+\nonumber\\
&& \left[\frac{1}{\sin^2\theta}-\frac{a^2}{\Delta}\right] 
\partial_{\phi\phi}\Psi +\, 2s \left[\frac{a (r-M)}{\Delta} 
+ \frac{i \cos\theta}{\sin^2\theta}\right] \partial_\phi\Psi  \nonumber\\
&&- \left(s^2 \cot^2\theta - s \right) \Psi = 0,
\end{eqnarray}
where in addition to the previously defined quantities, $s$ refers to the 
``spin weight'' of the matter field. As remarked before, this evolution 
equation determines the dynamical behavior of matter fields in the spacetime 
of a Kerr black hole. The $s = 0$ version of this equation describes the 
evolution of a scalar field $\Psi$ in a black hole spacetime, which is the 
case of interest in this work. 

To remove the coordinate singularity at the horizon locations, we consider 
the above equations in a different coordinate system. We summarize below the 
ingoing Kerr coordinate system $({\tilde t},r,\theta, {\tilde \varphi})$ and 
also the Teukolsky equation in these so-called ``horizon penetrating'' 
coordinates. In ingoing Kerr coordinates, the Kerr metric is given by 
\begin{eqnarray}
\,ds^2=\left(1-\frac{2Mr}{\Sigma}\right)\,d{\tilde t}^2-
\left(1+\frac{2Mr}{\Sigma}\right)\,dr^2-\Sigma\,d\theta^2\nonumber \\ 
-
\sin^2\theta\left(r^2+a^2+\frac{2Ma^2r}{\Sigma}\sin^2\theta\right)
\,d{\tilde \varphi}^2-\frac{4Mr}{\Sigma}\,d{\tilde t}\,dr\nonumber \\
+
\frac{4Mra}{\Sigma}\sin^2\theta\,d{\tilde t}\,d{\tilde \varphi}+
2a\sin^2\theta\left(1+\frac{2Mr}{\Sigma}\right)\,dr\,d{\tilde \varphi}\, .
\end{eqnarray}
These coordinates are related to the Boyer-Lindquist coordinates through 
the transformations ${\tilde \varphi}=\varphi+\int a\Delta^{-1}\,dr$ and 
${\tilde t}=t-r+r_*$, where the ``tortoise'' radial coordinate 
$r_*=\int(r^2+a^2)\Delta^{-1}\,dr$. This system does not suffer from any 
pathologies at the horizon locations and is therefore well-suited for analyzing 
fields both in the exterior and interior spacetimes of a rotating black hole.

\begin{center}
  \begin{figure}[h]
  \includegraphics[width=.45\textwidth ]{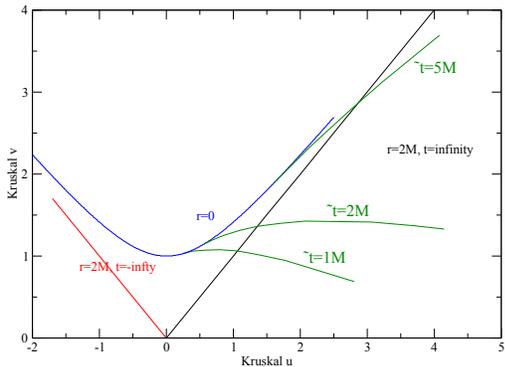}
  \caption{Kruskal diagram of the Eddington Finkelstein coordinates. 
  The depicted curves show the behavior of the constant-$\tilde t$ slices over the black hole 
  exterior and interior regions. }
  \label{fig:coords}
  \end{figure}
  \end{center}

It is important to point out how the physical meaning of these 
$({\tilde t},r,\theta, {\tilde \varphi})$ coordinates changes as one 
approaches and crosses the horizon from the exterior region of the black hole 
spacetime into the interior. To illustrate this, we switch to the Schwarzschild 
case for simplicity, by setting the Kerr parameter $a = 0$. In that case, the 
ingoing Kerr coordinates change to the more familiar Eddington-Finkelstein 
coordinates, and yet retain their qualitative behavior at the (outer) horizon. 
In Figure~\ref{fig:coords} we show the Kruskal diagram of the 
$\tilde t =$ constant slices over the entire exterior and interior regions. 
It is clear that the slices stay well behaved as they cross the horizon
and even stay spacelike throughout. However, this is not the case for the $r=$ 
constant slices. The constant-$r$ slices (note that for these slices 
$ds^2=\left(1-\frac{2M}{r}\right)\,d{\tilde t}^2$ if $d\theta=d{\tilde\varphi}=0$) 
are timelike outside the horizon and switch to becoming spacelike in the black hole 
interior region. 

The fact that the constant-$r$ slices are not timelike everywhere poses a 
challenge in how we interpret the time evolution results of a physical field. 
Typically, a scalar field's  values $\Psi({\tilde t}, r, \theta, {\tilde \varphi})$ 
at a fixed $(r, \theta, {\tilde \varphi})$ location would be interpreted as 
the time-series result of a detector sampling the field at that spatial point. 
The detector is, of course, a physical object and therefore it must have a timelike 
worldline through the spacetime. This works just as expected in the exterior region 
of the black hole spacetime. However, in the black hole interior, clearly the detector 
cannot be located at a constant value of $r$ because that would translate to a 
spacelike worldline. Any physical detector must take a timelike path which results 
in a decrease in its $r$ coordinate value as $\tilde t$ advances. Once the $r=0$ 
singularity is reached (which occurs in a finite proper time), any physical meaning 
associated to these mathematical quantities is simply lost. Thus, caution must 
be taken in interpreting any results from a $\tilde t$ evolution in the black hole 
interior region. 

Finally, it is useful to note that at the horizon location at late times, 
the $\tilde t$ variable is essentially the null variable $v = t + r_*$. 

%\begin{widetext}
%\end{widetext}

The Teukolsky equation for the scalar field $\Psi$ in these ingoing
Kerr coordinates can be derived using a rescaling of the Kinnersley 
tetrad~\cite{ptc}. It is given by 
\begin{eqnarray}
&&
(\Sigma + 2Mr){{\partial^2 \Psi}\over
{\partial \tilde t^2}} - \Delta {{\partial^2 \Psi}\over
{\partial r^2}} - 2 (r - M){{\partial \Psi}\over
{\partial r}} 
\nonumber\\
&&
-{{1}\over {\sin \theta}}{{\partial}\over {\partial \theta}} \left (   
\sin \theta {{\partial \Psi}\over {\partial \theta}}\right ) -{{1}\over
{\sin^2 \theta}}{{\partial^2 \Psi}\over {\partial \tilde \varphi^2}}  
-4Mr{{\partial^2 \Psi}\over {\partial \tilde t \partial r}}
\nonumber\\
&&
-2a {{\partial^2 \psi}\over {\partial r\partial \tilde \varphi}}
- 2{M} {{\partial \psi}\over {\partial \tilde t}} = 0\, ,
\end{eqnarray}
and is well-behaved at the horizon locations, and therefore can be used to 
safely evolve data across it. 

The last ingredient that goes into the setup of our coordinate system is 
hyperboloidal compactification as developed by Zengino\v{g}lu~\cite{anil}. 
To do this, we define a compactified coordinate system $(\tau,\rho,\theta,{\tilde\varphi})$ 
by
\begin{equation}
\tau = {\tilde{t}} - {r}^2/({r}+S) + 4 \ln [S/({r}+S)]
\end{equation}
and
\begin{equation}
\rho ={r}/[1+{r}/S]
\end{equation}
where a free parameter $S$ controls both the domain and also the foliation.  
Note that $\rho\in [0,S)$ maps $r\in [0,\infty )$ and is therefore a one-to-one 
compactifying coordinate. A Penrose diagram of the slices defined by these 
coordinates in the Kerr spacetime context can be found in Ref.~\cite{cauchy}. 
We do not show the final form of the Teukolsky master equation in these 
compactified coordinates because of the lengthy nature of the expression and 
the fact that it is not particularly illuminating. We simply refer the reader 
to the recent relevant research literature~\cite{hyper,harms} wherein additional 
details may be found.   

The computational grid is defined as a uniform grid over the compactified $\rho$ 
coordinate. As pointed out earlier, this allows us to access null infinity 
directly on the computational grid ($\rho = S$ maps to $\scri^+$).  Moreover, the 
compactification offers a ``clean'' solution to the so-called ``outer boundary problem'' 
in numerical relativity. Typical boundary conditions used in the research community 
lead to spurious wave reflections from the edge of the computational grid. However, 
with the approach of hyperboloidal compactification, one is able to extend the 
computational domain to infinity, making it possible to completely eliminate any such 
reflections~\cite{hyper}. In addition, the compactification allows us to employ a 
very dense computational grid (typically, $S\sim 20$) which results in highly accurate 
numerical results. Those details are provided in the next subsection. 

Once again, it is useful to point out that at the horizon locations the $\tau$ variable 
is essentially the same as the null variable $v$. 

\subsection{Computational Methodology}

The numerical approach used to solve the Teukolsky equation in the compactified 
ingoing Kerr coordinate system is very similar to the one presented in our earlier 
work~\cite{hyper}. We simply outline the main steps here and refer the reader to 
that reference for additional details. We begin by taking advantage of Kerr 
spacetime's axisymmetry and separating out the $\tilde\varphi$ dependence of the 
system using an $\exp(im{\tilde\varphi})$ form for the scalar field $\Psi$. This 
transforms the original (3+1)D equation into a system of (2+1)D equations. In this 
work we restrict ourselves to axisymmetric fields only, and therefore we set $m=0$ 
throughout. Next, we cast the equations into a first-order hyperbolic partial differential 
equation form, by defining a new ``momentum'' field that is related to the derivative 
of the scalar field $\Psi$. Finally, we implement a time-explicit,  two-step 
Richtmeyer-Lax-Wendroff, second-order\footnote{In fact, the angular differentiation 
(the $\theta$-derivatives) are implemented using a higher-order numerical stencil. 
This was deemed to be necessary to keep the {\em truncation} error at sufficiently 
low levels. The temporal and the radial direction related operations are second-order 
and such a mixed approach yields sufficiently good results.} finite-difference 
evolution scheme\footnote{In recent work~\cite{weno} we have developed a fifth-order 
WENO finite-difference scheme with third-order Shu-Osher explicit time-stepping. 
This new approach yields the same results.} This numerical method is stable, and 
converges to the expected second-order accuracy~\cite{hyper}.       

It is worth commenting on the fact that numerical computations are rather  
challenging in the context of studying the late-time tails. As remarked before, 
these computations must be long duration because the observed field initially exhibits 
an exponentially decaying oscillatory behavior, i.e. quasi-normal ringing. Only much 
later, once the exponential decay has made these modes subdominant, does the field   
transition over to a power-law tail. Moreover, there are often intermediate tails~\cite{grg}, 
that do not necessarily have the true late-time asymptotic rates that we are interested 
in here. These intermediate tails decay faster than the asymptotic rate, but may 
have dominant amplitudes for a period of time. We must evolve longer than these 
intermediate tails last in order to obtain the tail solution with the true asymptotic 
decay rate. 

In addition, each of the field's spherical harmonic multipoles $Y_{\ell m}$ has its 
own decay rate (that is proportional to $\ell$). Thus, at late times we obtain 
numerical data in which different multipoles may have widely ranging amplitudes 
(typically 20 -- 30 orders of magnitude apart!). It is thus important for the numerical 
solution to have high grid density in order to reduce the {\em truncation} errors 
to very low levels. In addition, due to the very large range of amplitudes involved, 
these computations also require high-precision floating-point numerics that allow us 
to reduce {\em round-off} error that can otherwise easily overwhelm the fast decaying 
multipoles. In particular, we satisfy this requirement by using {\em octal}-precision 
numerics (256-bit or $\sim$60 decimal digits). This keeps the round-off error in our 
computations at acceptably low levels. 

Finally, to complete these long duration, high-accuracy and high-precision computations 
in a reasonable time-frame we make extensive use GPGPU-based parallel computing. For 
additional details on implementation of such intensive computations on a parallel GPU 
architecture, we refer the reader to our earlier work on the subject~\cite{jsc}. 

\section{Numerical Results}

In this section, we present the results generated from the numerical solution of the 
scalar Teukolsky equation in compactified, ingoing Kerr coordinates. As outlined in the 
previous sections, our approach allows us to evolve data through the horizon and we are 
thus able to study the late-time behavior of physical fields in both the black hole 
exterior and interior regions. Our emphasis in this work is the late-time, power-law tails 
behavior of different spherical harmonic multipoles $Y_{\ell m}$ of a scalar field in 
both Schwarzschild and Kerr black hole spacetimes. In addition, we restrict ourselves to 
the axisymmetric case only in this work. 

Throughout this work, we choose the initial data for the scalar field to be a  
Gaussian distribution localized at $\rho = 8M$. The angular distribution is an axisymmetric 
spherical harmonic of multipole $\ell$. We include results for both {\em compact} (a truncated 
Gaussian with width $0.1M$) and {\em non-compact} (a wide Gaussian) initial data. For the Kerr 
cases, the value of $a/M = 0.8$ and the inner boundary with a Neumann boundary conditions is located 
at the inner horizon. In the Schwarzschild cases, we place the inner boundary at $\rho=0.05M$. 
Our results are completely insensitive to the location and type of boundary condition. 
The outer boundary of the computational domain is located at null infinity. 

\subsection{Scalar Fields in Schwarzschild Black Hole Spacetime}

We begin with the results for the non-rotating Schwarzschild black hole spacetime case. 
We present our power-law tails data for the exterior spacetime (horizon, timelike and 
null infinity) first, followed by the same in the interior region for a number of 
different multipoles $\ell$. We show results for both types of initial data, i.e. with 
and without compact support. 

\subsubsection{Exterior Spacetime}

The expected outcome for the exterior spacetime case is the well-established power-law tail 
$\tau^{-2\ell-3}$ \cite{price} which is valid at the horizon and timelike infinity, while 
the expression changes to $\tau^{-\ell-2}$ at null infinity~\cite{price}. In these expressions,
$\ell$ is the multipole of the scalar field under consideration, and recall that at the 
horizon location $\tau$ is essentially the null coordinate $v$. 

The data depicted in Table~\ref{schw_rates}  
shows that our approach reproduces these well-established tails and therefore, 
serves as an excellent check on our methodology. It is worth noting that we tested our 
implementation for all multipoles up to $\ell = 10$ with excellent agreement with the 
power-law tail expression~\cite{price}. This was found to be the case for both types of 
initial data (compact and non-compact). 

\begin{table}[h]
  \begin{center}
      \begin{tabular}{|c||c|c|c|}
        \hline
        $\ell \backslash \rho$ & Horizon & $\rho$ = 3.0 & Infinity $\scri^+$ \\
        \hline\hline 
        2                       & -7 & -7 &-4   \\
        3                       & -9 & -9 &-5   \\
        4                       &-11 &-11 &-6   \\
        5                       &-13 &-13 &-7   \\
        6                       &-15 &-15 &-8   \\
        8                       &-19 &-19 &-10  \\
        \hline
   \end{tabular}
  \end{center}
\caption{Asymptotic late-time scalar field tails in exterior Schwarzschild space-time
for several multipoles. The power-laws agree precisely with the expected law~\cite{price}.}
  \label{schw_rates}
\end{table}

\subsubsection{Interior Spacetime}

\begin{figure}[h]
  \begin{center}
  \includegraphics[width=.40\textwidth ]{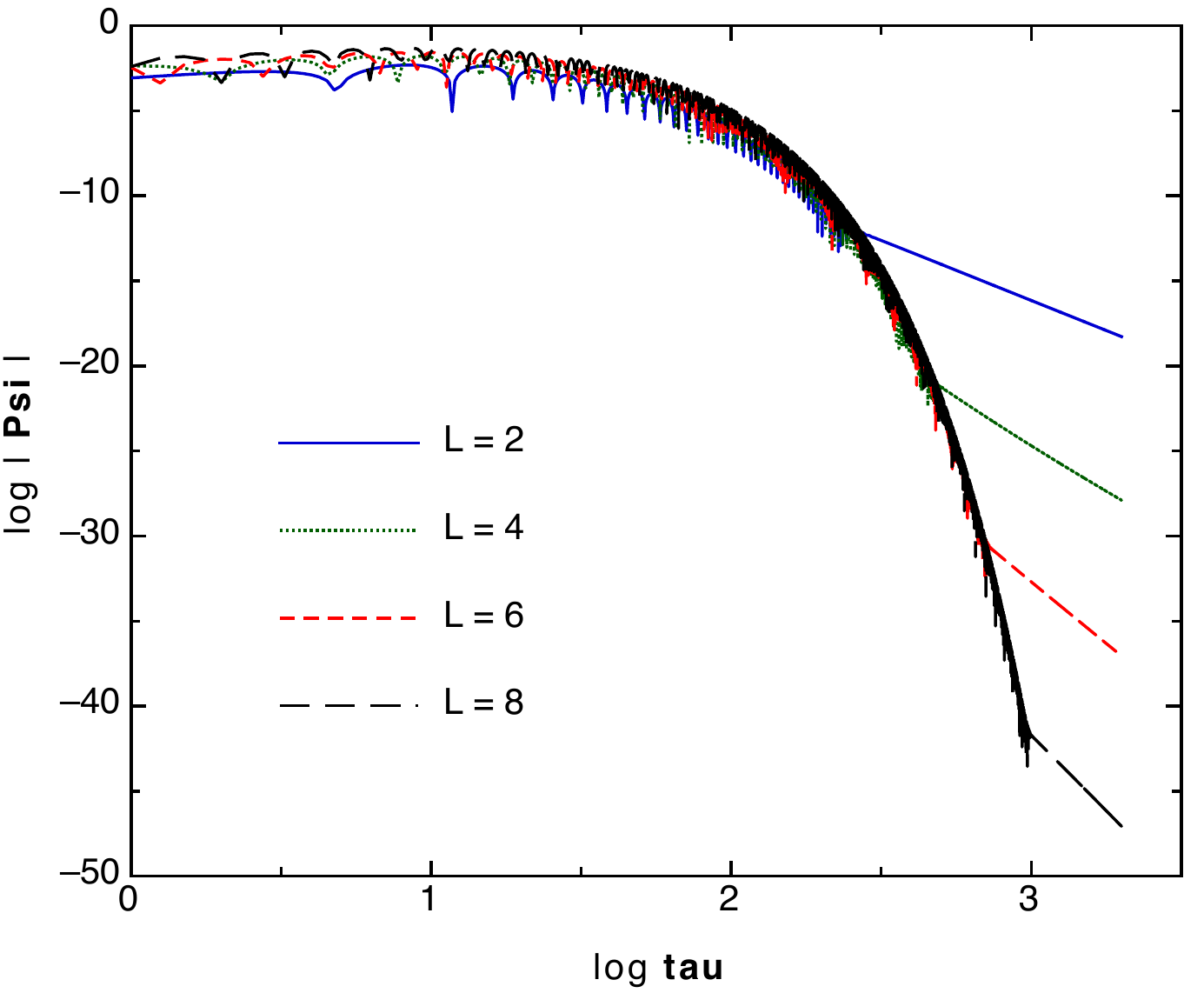}
  \caption{ 
  Late-time scalar field tails for different multipoles $\ell$ at $\rho = 1$ for the case of  
  Schwarzschild black hole interior. The power-laws agree with the expression $\tau^{-2\ell-3}$.}
  \label{fig:schw_0}
  \end{center}
\end{figure}

Here we report on some new results in the case of the interior region of the Schwarzschild 
spacetime. As cautioned before, it is not meaningful to sample the field at a fixed $\rho$ 
location, because that implies that the detector is on a spacelike worldline. However, 
nonetheless, if we proceed with that type of data sampling anyhow, it is interesting that 
the field exhibits late-time power-law decays in the $\tau$ coordinate, that generically 
satisfy the usual law~\cite{price}. This can be seen in Figure~\ref{fig:schw_0} for a number of different 
$\ell$ multipoles. For the high-$\ell$ cases it is clear that high-precision numerics are 
necessary, since the field's amplitude falls very rapidly due to the fast decay rate.  

For a more physically meaningful detector, we remarked before that a timelike worldline 
dictates that the detector must continuously fall towards $\rho=0$, i.e. the radial coordinate 
must decrease in value monotonically. To understand what type of signal such a detector 
would record, we plot the snapshots of the field as a function of $\rho$ at different 
values of $\tau$. 

\begin{figure}[h]
  \begin{center}
  \includegraphics[width=.40\textwidth ]{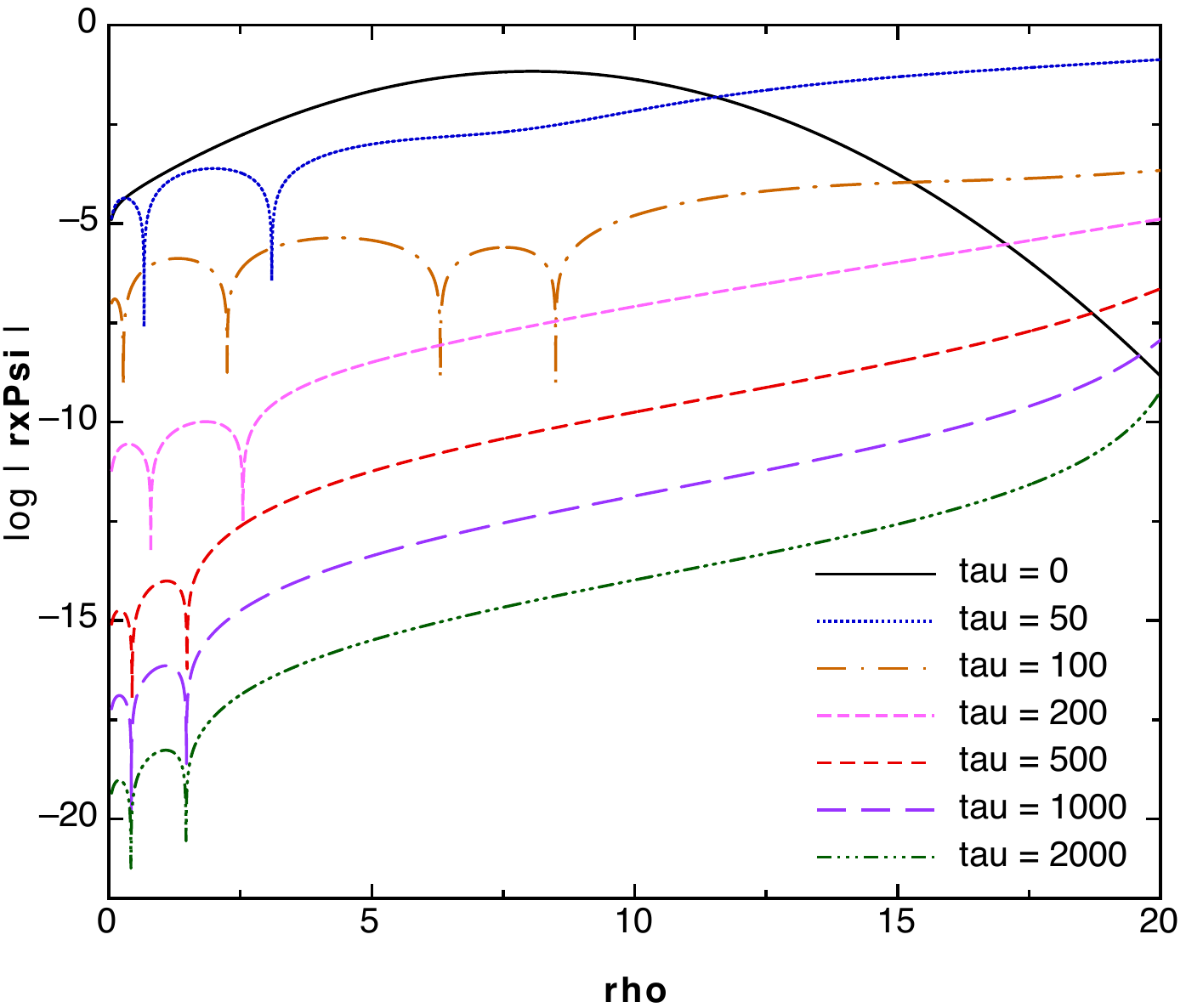}
  \caption{ 
  Late-time scalar field radial $\rho$ profile for multipole $\ell=2$ for  
  a Schwarzschild black hole. Note that even at late-stages, the field exhibits 
  oscillatory behavior but only in the black hole interior region.}
  \label{fig:schw_rho_1}
  \end{center}
\end{figure}

\begin{figure}[h]
  \begin{center}
  \includegraphics[width=.40\textwidth ]{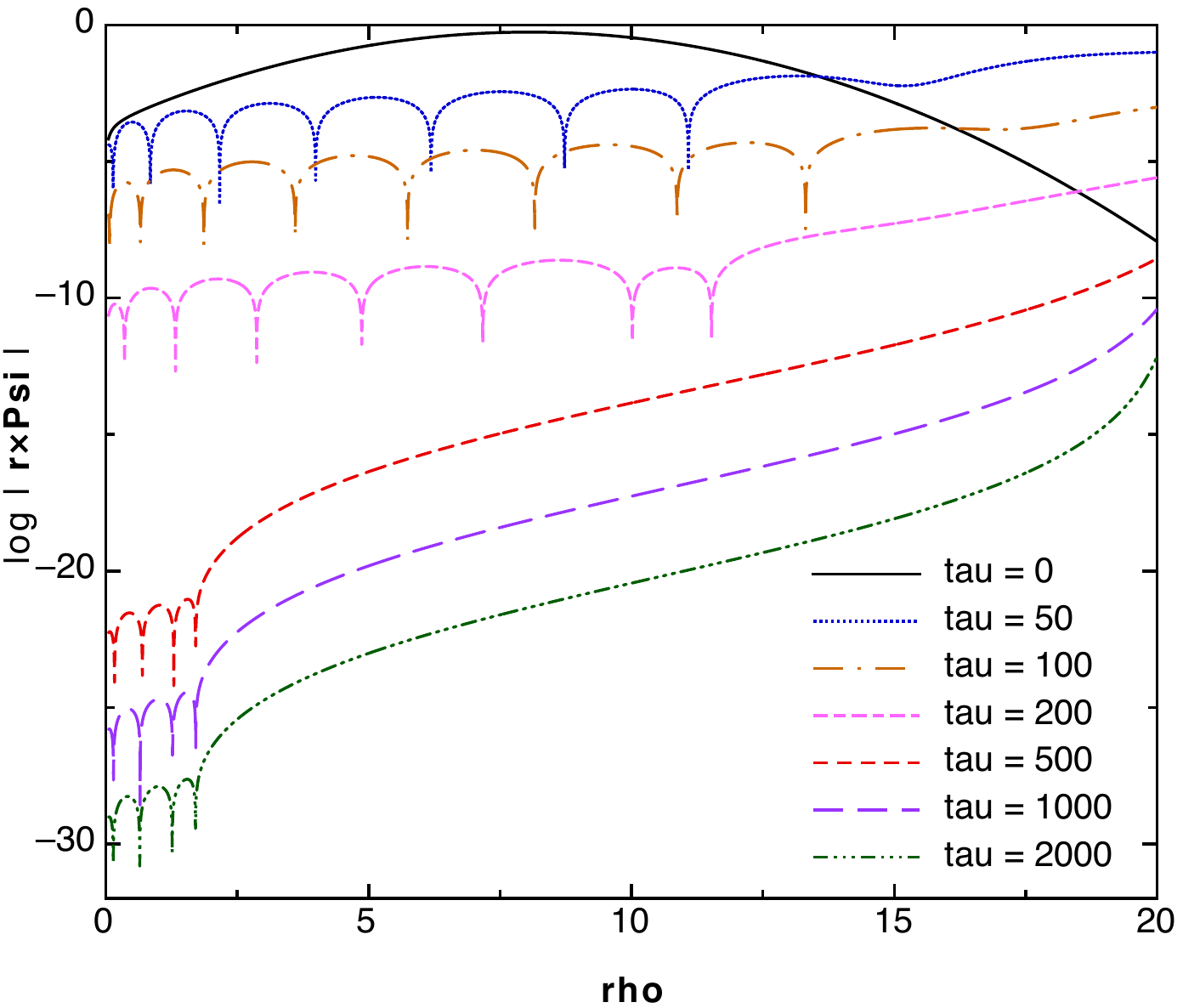}
  \caption{ 
  Late-time scalar field radial $\rho$ profile for multipole $\ell=4$ for  
  a Schwarzschild black hole. Note that even at late-stages the field exhibits 
  oscillatory behavior but only in the black hole interior region. 
  The number of cycles is proportional to $\ell$.}
  \label{fig:schw_rho_2}
  \end{center}
\end{figure}

The case of non-compact initial data, is shown in Figures~\ref{fig:schw_rho_1}, \ref{fig:schw_rho_2}
\footnote{Animations based on the dynamics of the field's radial profile may be found at 
this URL: \url{https://www.youtube.com/channel/UCVNLqTQx1O2sbwc-4wsE_Gw}}. 
At early times, we observe many dynamical features as the field evolves through 
the quasi-normal ringing stage and finally ``settles'' to more quiescent tail state. In the 
late-stage tails regime, the field appears to maintain its $\rho$ profile and simply fall in 
amplitude determined by the usual power-law~\cite{price}. However, we observe a number of interesting features in the 
interior region of the spacetime. In particular, there are $\rho$ ``locations'' where the field 
switches sign  (this does not happen in the exterior region). And, in fact, the number of these 
zero-crossings or ``nodes'' increases proportionately with the multipole $\ell$ value. This suggests 
that a physical detector would observe a finite number of oscillatory cycles in the field 
{\em after} crossing into the black hole interior before the detector hits the central black 
hole singularity. The higher the value of $\ell$, the larger the number of cycles the detector 
would record. 

\begin{figure}[h]
  \begin{center}
  \includegraphics[width=.40\textwidth ]{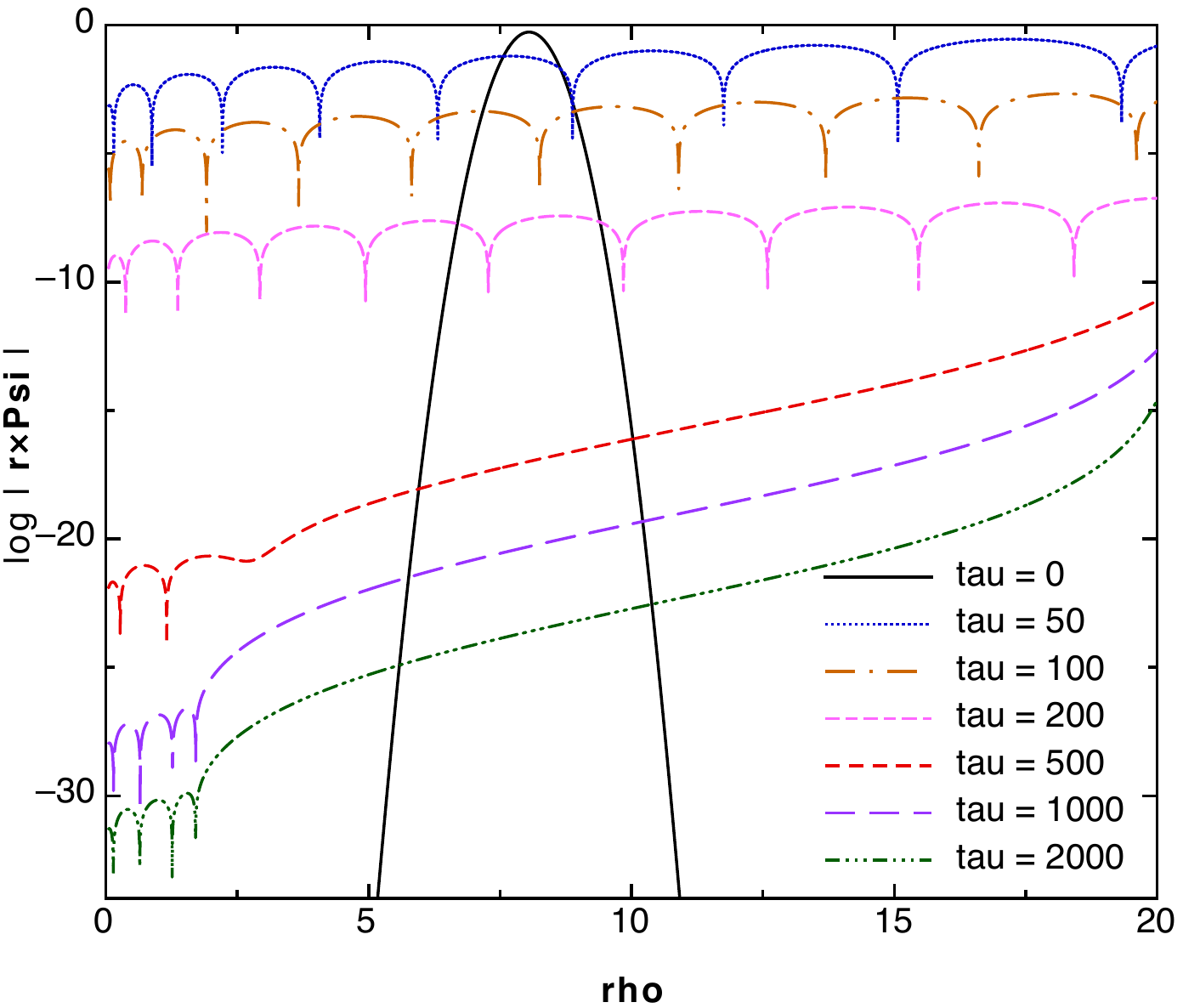}
  \caption{ 
  Late-time scalar field radial $\rho$ profile for multipole $\ell=4$ for  
  a Schwarzschild black hole. The initial field has compact support. At late-stages 
  the field exhibits oscillatory behavior but only in the black hole interior region. 
  The number of cycles is proportional to $\ell$.}
  \label{fig:schw_cpt_rho}
  \end{center}
\end{figure}

The exact same features also appear in the case of the compact initial data. For 
that reason, we only document results from one multipole alone, i.e. $\ell = 4$ in 
Figure~\ref{fig:schw_cpt_rho}. The field behaves similar to the non-compact case and the 
same number and type of zero-crossings are observed in the interior region. 

At a mathematical level, it is not difficult to uncover the detailed explanation for 
these oscillations in the interior. Here we derive {\em analytic} solutions for the 
location of these late-time nodes found within the Schwarzschild black hole horizon. We 
begin with the Teukolsky equation for the scalar field $\Psi$ (Eqn. 4) and consider the 
Schwarzschild case by setting $a = 0$. 

Next, we assume separability~\cite{RHPthesis} of the angular, radial and temporal functions 
such that $\Psi = Y_{\ell m} (\theta, \tilde \varphi) R(r) T(\tilde t)$ where the $Y_{\ell m}$ 
functions are the well-known spherical harmonics. Thus, the radial and temporal part of 
the separated differential equation take the form
\begin{eqnarray}
&&
\ell(\ell+1) = -\frac{1}{T} \left((r^2+2Mr){{\partial^2 T}\over {\partial \tilde t^2}}- 
2M{{\partial T}\over {\partial \tilde t}}\right)\nonumber\\
&&
+\frac{1}{R} \left( (r^2-2Mr){{\partial^2 R}\over {\partial r^2}}  + 
2(r-M){{\partial R}\over {\partial r}} \right)\nonumber\\
&&
+\frac{1}{RT} 4Mr {{\partial T}\over {\partial \tilde t}}{{\partial R}\over{\partial \tilde r}}. 
\end{eqnarray} 
The last term in the equation above prevents clear separability between the radial and temporal 
dependences. However, motivated by our numerical results, if we let the behavior of the temporal 
component to be a power law decay, i.e. $T(\tilde t) \approx \tilde t^{-n}$, then in the limit of 
$\tilde t \rightarrow \infty$ the above equation reduces to 
\begin{equation}
(r^2-2Mr){{\partial^2 R}\over {\partial r^2}} + 2(r-M){{\partial R}\over {\partial \tilde r}} 
=\ell(\ell+1)R(r).
\end{equation}
This differential equation takes the form of the well documented Sturm--Liouville problem, with the 
well-known solutions
\begin{equation}
    R(r) = C_{1} P_{\ell}(r/M-1)+C_{2} Q_{\ell}(r/M-1)
\end{equation}
where $P$ and $Q$ are the Legendre polynomials of the first and second kind respectively\footnote{
We note that this solution was also obtained by Ori in Ref.~\cite{ori98}.}. For the 
solution above to exhibit the correct behavior at the horizon, we must set $C_2 = 0$ and therefore 
\begin{equation}
    R(r) \propto P_{\ell}(r/M-1)
\end{equation}
The roots of $R(r)$ above are in excellent agreement with the location of the nodes we present in 
Figs.~\ref{fig:schw_rho_1}, \ref{fig:schw_rho_2} and \ref{fig:schw_cpt_rho}. Thus, we see that 
the oscillations as observed by an infalling physical detector come from the oscillations in the 
Legendre polynomials that describe the state of the scalar field on the interior.

\subsection{Scalar Fields in Kerr Black Hole Spacetime}

In this subsection we present our results for the rotating Kerr black hole case. Once again 
we present our power-law tails data for the exterior spacetime (horizon, timelike and null 
infinity) first, followed by the same in the interior region for a number of different 
multipoles. 

\subsubsection{Exterior Spacetime}

Interestingly, only recently has an understanding of the nature of the late-time 
power-law tails in Kerr spacetime been fully uncovered~\cite{grg,BK}. The late-time 
decay rate expressions for the scalar field case are given by $\tau^n$ where 
\begin{equation}  \label{eq:rates}\quad
n = \left\{ \begin{array}{ll}
-(\ell'+\ell + 3) & \mathrm{for}\quad \ell'=0, 1 \\
-(\ell'+\ell + 1) & \mathrm{otherwise}\quad \end{array}\right.\\
\end{equation}
on the horizon and timelike infinity, and 
\begin{equation} \label{eq:rates2}\quad
n^{\scri^+} = \left\{ \begin{array}{ll}
-\ell' & \mathrm{for}\quad \ell\leq \ell'-2  \\
-(\ell+2) & \mathrm{for} \quad \ell\geq \ell' \end{array}\right.
\end{equation}
at null infinity. Here $\ell'$ refers to the initial field multipole and $\ell$ is the 
projected multipole of the full late-time field under consideration. In the Schwarzschild 
case, these are always the same, i.e. $\ell' = \ell$. However, since Kerr spacetime is 
not spherically symmetric, a pure $\ell'$ multipole does not stay pure as it is evolved, 
i.e. other $\ell$ multipoles are ``excited''. In general, even-valued $\ell'$ modes 
only excite even $\ell$, while the odd-valued ones excite only the odd multipoles. 
This is because Kerr spacetime still retains a reflection symmetry about the 
equatorial plane.  

The above expressions were obtained by carefully studying the ``inter-mode coupling'' 
effects that are present in Kerr space-time due to frame-dragging~\cite{BK}. Note 
that these expressions above are only for the axisymmetric multipoles. In our work, 
we only study the full field throughout, which implies we study the late-time dominant 
multipole $\ell = 0$ for the even $\ell'$ case and $\ell = 1$ for the odd case. 
At late times, clearly these will dominant all other multipoles (they exhibit the slowest decay) 
and therefore, dominate in the full field. Thus, the above expressions simplify to $\tau^{-\ell'-1}$ 
for even $\ell' > 0$ and $\tau^{-\ell'-2}$ for odd $\ell' > 1$ at the horizon and timelike 
infinity. At null infinity they reduce to $\tau^{-\ell'}$ for $\ell' > 1$. In this 
section, we verify these expressions using our numerical data. 

Once again, Table~\ref{kerr_rates} shows  
that our approach works correctly and the equations~\ref{eq:rates}, \ref{eq:rates2} 
are verified. We tested our implementation for all multipoles up to $\ell' = 8$ with 
excellent agreement. This is the case for both types of initial data. 

\begin{table}[h]
  \begin{center}
      \begin{tabular}{|c||c|c|c|}
        \hline
        $\ell' \backslash \rho$ & Horizon & $\rho$ = 3.0 & Infinity $\scri^+$ \\
        \hline\hline 
        2                       &-3 &-3 &-2   \\
        4                       &-5 &-5 &-4   \\   
        6                       &-7 &-7 &-6   \\
        8                       &-9 &-9 &-8   \\
        \hline
   \end{tabular}
  \end{center}
\caption{Asymptotic late-time scalar field tails in exterior Kerr space-time
for several multipoles. The power-laws agree with equations~\ref{eq:rates}, \ref{eq:rates2}.}
  \label{kerr_rates}
\end{table}

\subsubsection{Interior Spacetime}

\begin{figure}[h]
  \begin{center}
  \includegraphics[width=.40\textwidth ]{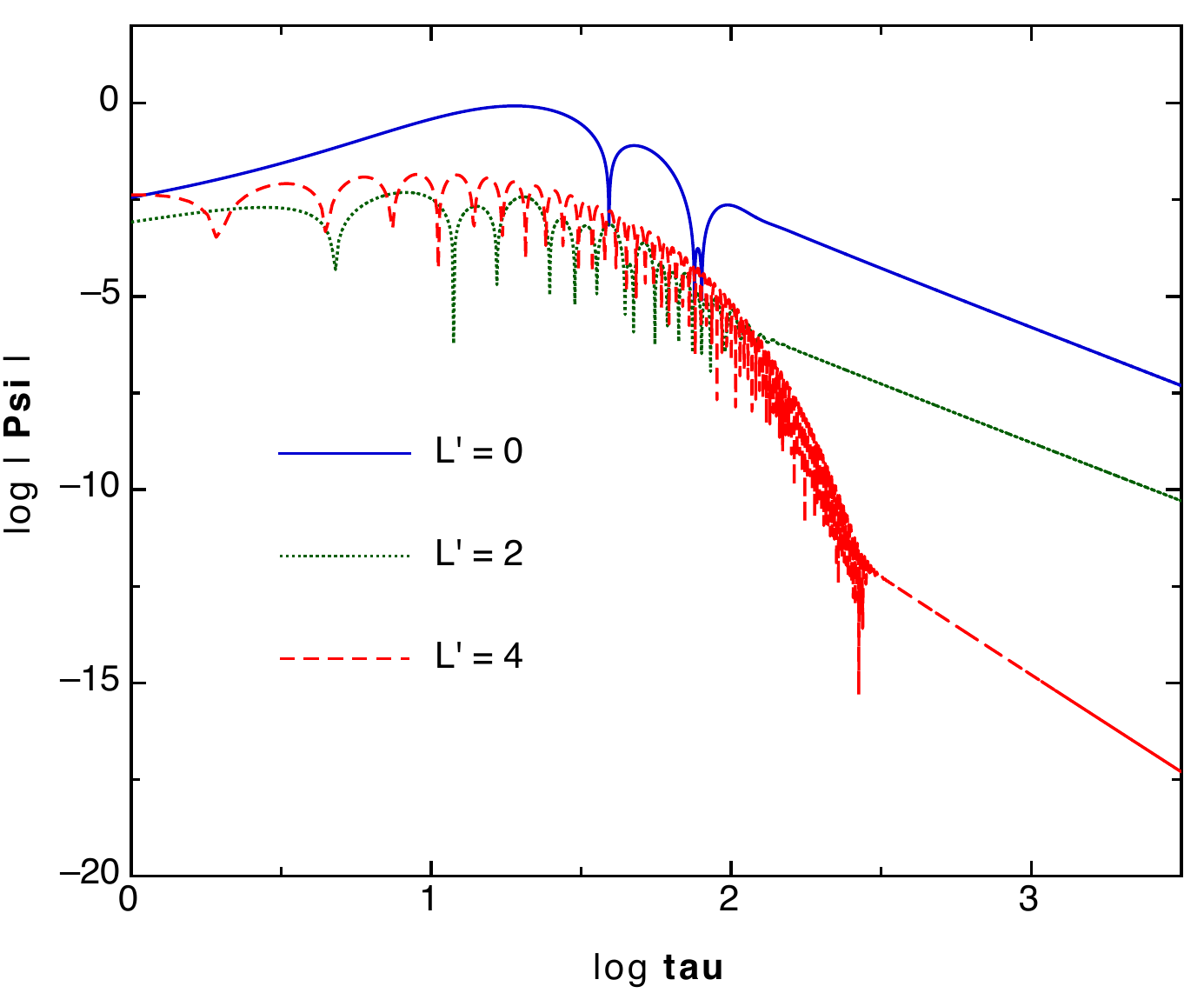}
  \caption{ 
  Late-time scalar field tails for different multipoles $\ell'$ at $\rho = 1$ of 
  a Kerr black hole interior. The power-laws agree with the equation~\ref{eq:rates}.}
  \label{fig:kerr_0}
  \end{center}
\end{figure}

Our new results in the context of the interior region of the Kerr spacetime are similar 
to the Schwarzschild case. If we proceed with sampling the field data at a fixed $\rho$ 
it exhibits late-time power-law decays in the $\tau$ coordinate, that generically 
satisfy the equivalent of the well-known tails law~\cite{price} for Kerr, equation~\ref{eq:rates}. This can be 
seen in Figure~\ref{fig:kerr_0} for a number of different $\ell'$ multipoles. 

\begin{figure}[h]
  \begin{center}
  \includegraphics[width=.40\textwidth ]{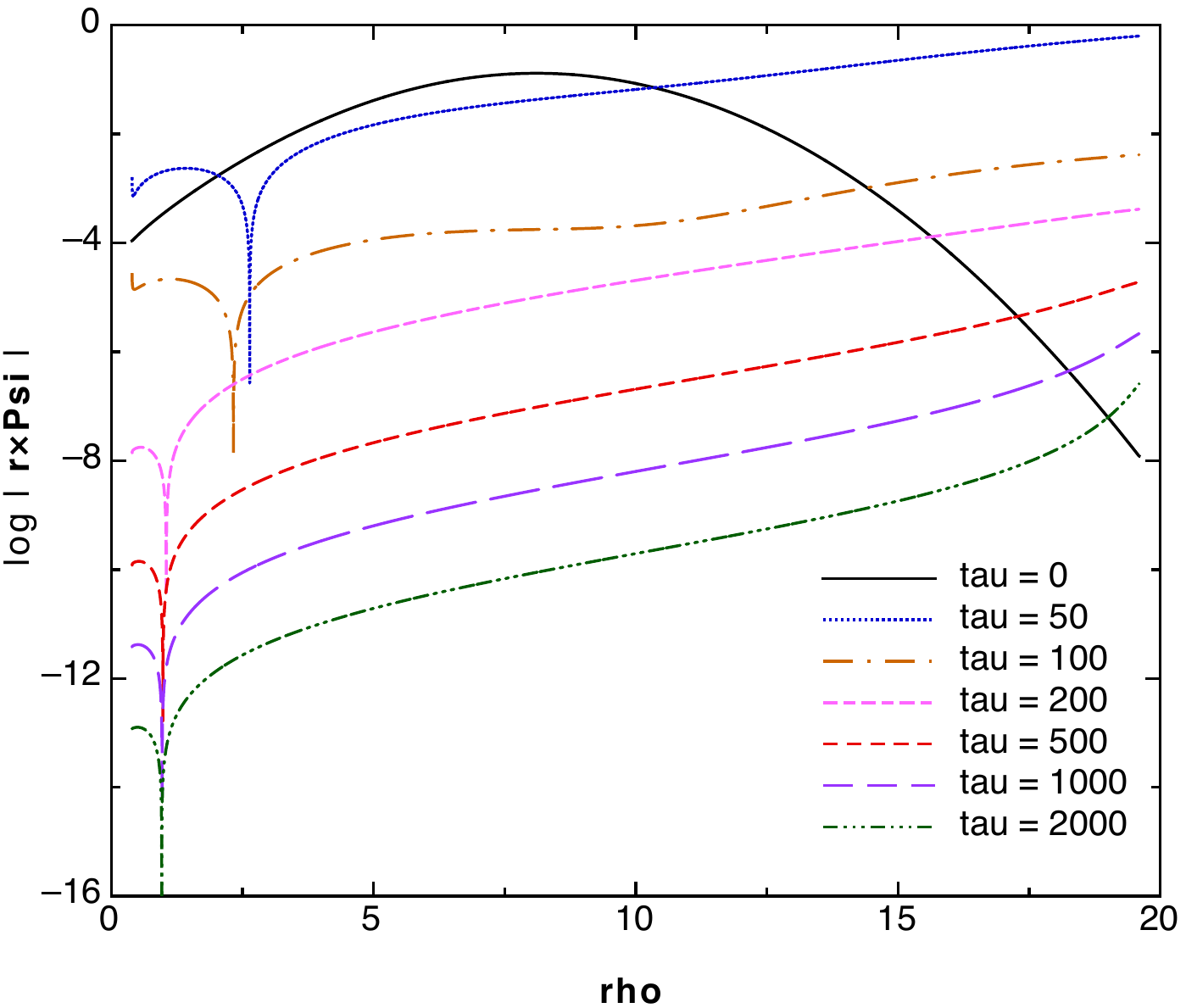}
  \caption{ 
  Late-time scalar field radial $\rho$ profile for multipole $\ell' = 1$ for  
  a Kerr black hole. Note that even at late-stages the field exhibits 
  oscillatory behavior but only in the black hole interior region. }
  \label{fig:kerr_rho_1}
  \end{center}
\end{figure}

\begin{figure}[h]
  \begin{center}
  \includegraphics[width=.40\textwidth ]{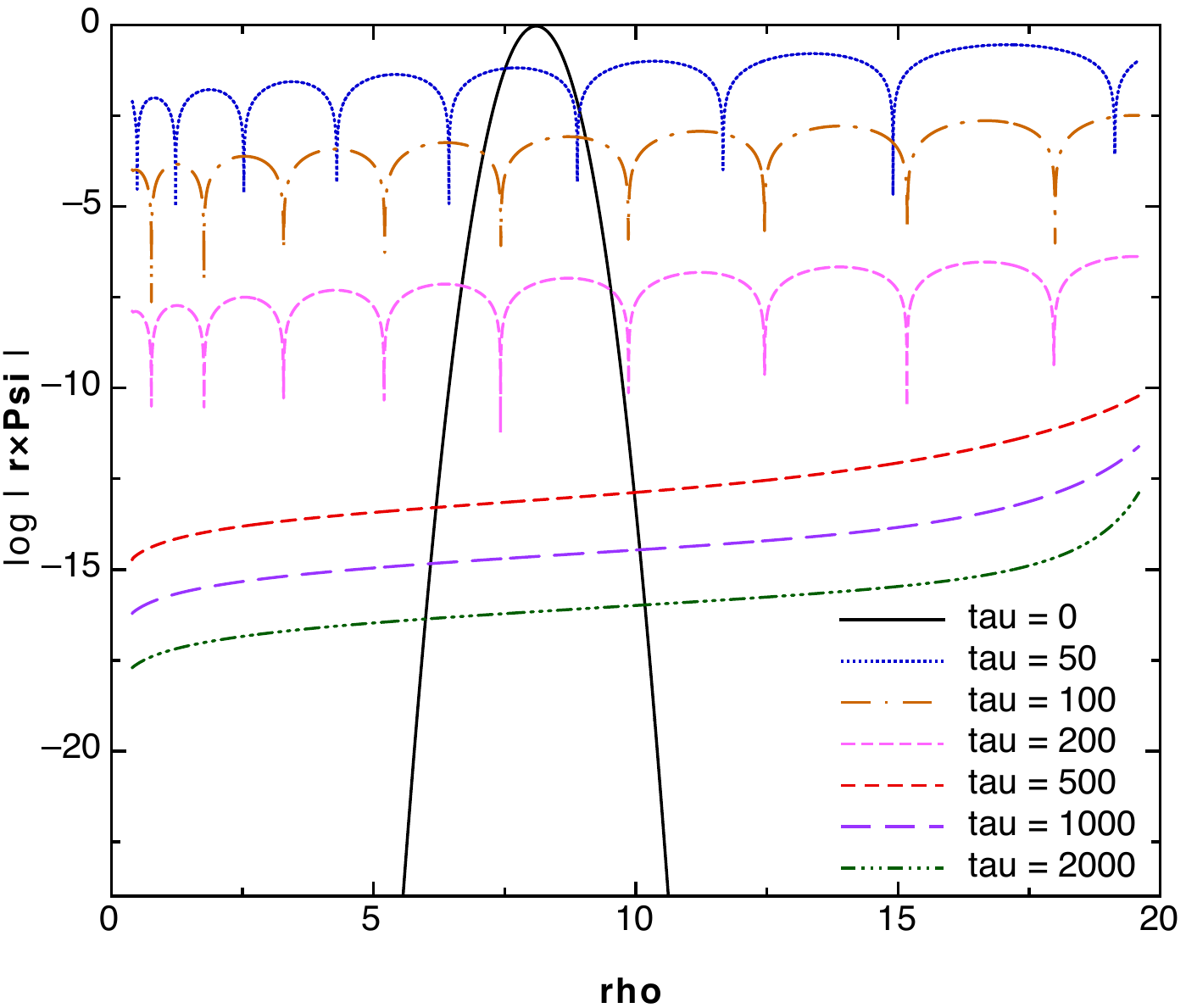}
  \caption{ 
  Late-time scalar field radial $\rho$ profile for multipole $\ell' = 4$ for  
  a Kerr black hole. There are no oscillations at late times because of the dominance 
  of the $\ell =0$ multipole.}
  \label{fig:kerr_rho_2}
  \end{center}
\end{figure}

For a more physically meaningful detector, one that must continuously fall towards $\rho=0$ 
once again we observe a finite number of oscillatory cycles due to the formation of 
zero-crossing nodes in the interior region. This can be seen in Figures~\ref{fig:kerr_rho_1}, 
\ref{fig:kerr_rho_2} that depict the multipole data as a function of $\rho$ for several 
moments in $\tau$. However, we do not see as many oscillatory cycles as observed in the 
Schwarzschild case. This is simply because of the dominance of the excited low-$\ell$ modes 
at late times that occurs in Kerr spacetime, but not in Schwarzschild. In fact, we observe 
precisely one zero-crossing in the odd-$\ell'$ case and none at all in the even-$\ell'$ case. 
This is in line with expectations, of course, since $\ell = 1$ is the lowest odd multipole 
that is excited, while $\ell = 0$ is the lowest one for the even case. We observe the exact 
same features in the case with compactly supported initial data. 

Finally, we also computed the late-stage, power-law tails of a scalar field of different 
initial multipoles at the Cauchy (inner) horizon of the black hole. Recall that at the 
horizon locations, the variable $\tau$ is essentially the null variable $v$. Therefore, 
these results may be compared with Ori's well-known work in a similar context~\cite{ori2}. 
They are in excellent agreement. 

\subsection{Scalar Fields in Extremal Kerr Black Hole Spacetime}

Recently~\cite{aretakis,gralla} it has been shown that massless scalar fields in extremal 
black hole spacetimes exhibit an ``asymptotic'' instability at the horizon. This arises 
in the form of an unbounded growth of (sufficiently) high-order transverse derivatives of 
the field. Heuristically, the origin of this instability is in the fact that the power-law 
tail's decay rate at the horizon is slower than at a ``nearby'' location on the exterior. 
This {\em only} occurs in the case of an extremal hole (note that our results in the previous 
section on non-extremal Kerr consistently yielded the exact same tail on the horizon as the 
timelike infinity case). The different decay rate at the horizon results in an unbounded 
growth in high-enough transverse derivatives. More precisely this instability can be seen 
as arising from a singular branch-point in the frequency-domain Green function~\cite{gralla}. 
Our numerical results in the context of extremal Kerr black holes are able to show strong 
indications of the development of such an instability. 

\begin{figure}[h]
  \begin{center}
  \includegraphics[width=.40\textwidth ]{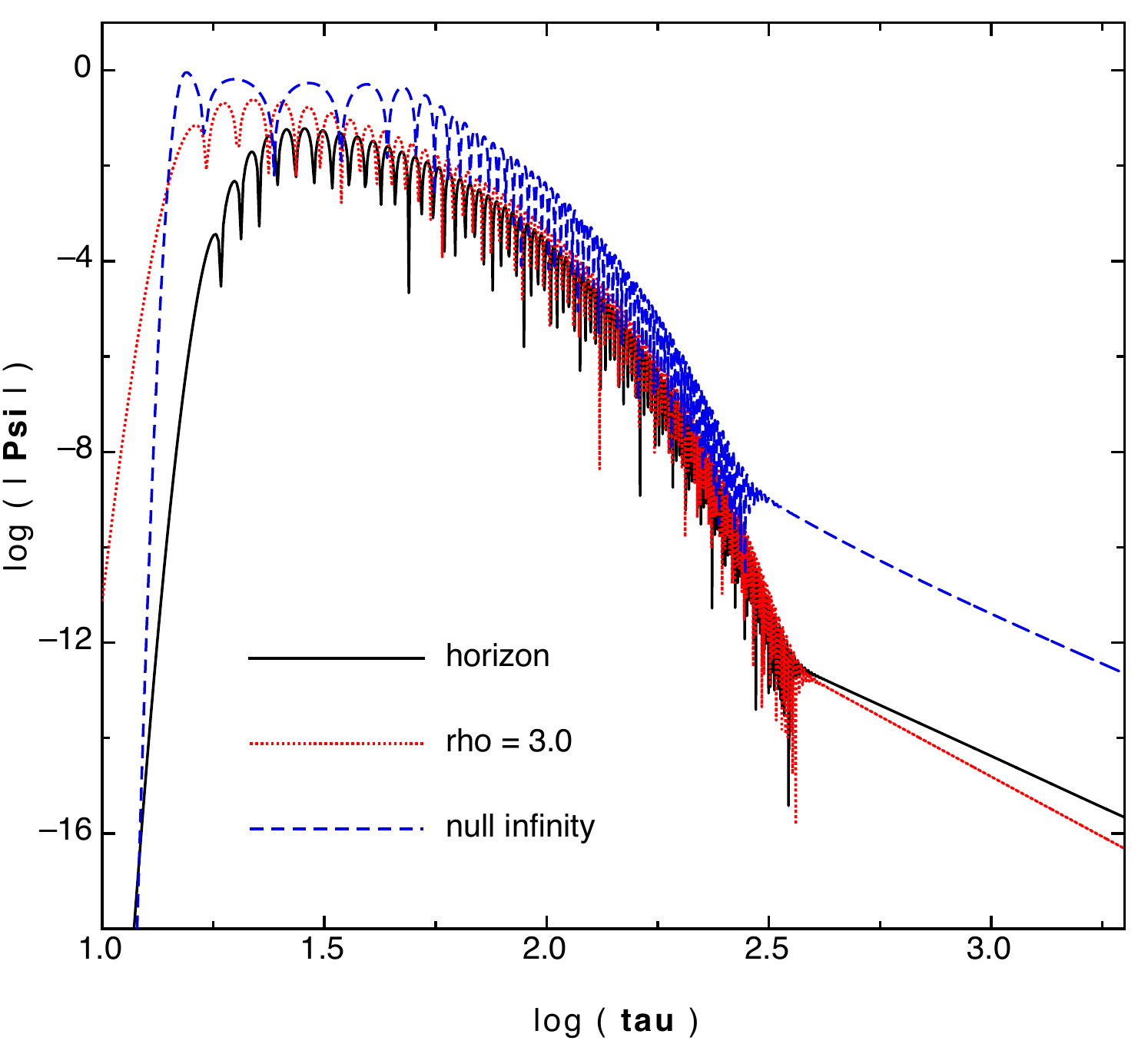}
  \caption{ 
  Late-time scalar field tails for $\ell' = 4$ at different locations (horizon, $\rho = 3.0$ 
  and null infinity) for an extremal Kerr black hole. The horizon decay rate is slower 
  than the $\rho = 3.0$ case, and matches the rate at null infinity.} 
  \label{fig:ekerr_2}
  \end{center}
\end{figure}

In Figure~\ref{fig:ekerr_2} we show the late-time tails for the 
extremal Kerr case for a sample multipole with compact initial data -- it is clear that 
the horizon decay rates are slower (they actually match the null infinity rates!) than the 
timelike infinity cases for the same multipole. 

\begin{figure}[h]
  \begin{center}
  \includegraphics[width=.40\textwidth ]{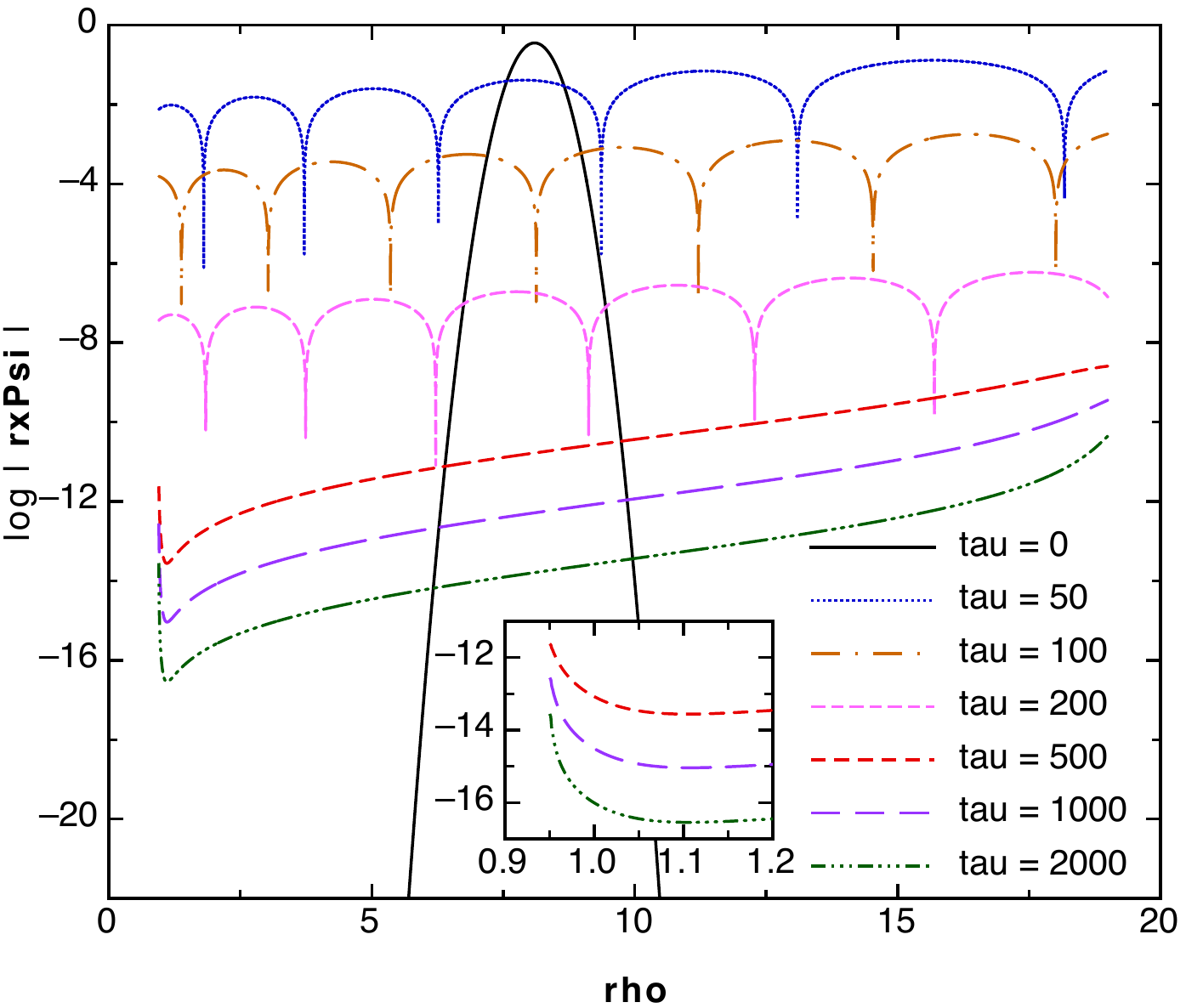}
  \caption{ 
  Late-time scalar field radial $\rho$ profile for multipole $\ell' = 3$ for  
  an extremal Kerr black hole. Note that the field's gradient exhibits unbounded growth 
  at the horizon.}
  \label{fig:ekerr_rho}
  \end{center}
\end{figure}

Moreover, if we study the radial ($\rho$) profile of the field as it evolves in time, 
it clearly appears to show the formation of an asymptotic ``discontinuity'' at the horizon. 
In Figure~\ref{fig:ekerr_rho} we can see that the gradient of the field as it evolves becomes 
steeper at the horizon, a strong indication of the transverse derivatives becoming unbounded, 
consistent with the expectations based on recent work~\cite{gralla}. A detailed study on this 
topic will be presented elsewhere~\cite{next}.

\section{Summary and Conclusions}

In this work, we performed a detailed study of the late-time tail behavior of scalar fields 
in black hole exterior and interior spacetimes. Both Schwarzschild and Kerr black holes were 
considered in this work. We numerically solved the scalar Teukolsky equation in compactified 
ingoing Kerr coordinates, and performed very long duration computations to obtain the true 
late-time asymptotic power-law decay behavior. We compared our numerical results with well  
known results in the research literature in the black hole exterior region (horizon, timelike 
and null infinity). The new results in this work pertain to the late-time behavior of scalar 
fields in the black hole interior. We found that an infalling detector would record a finite 
number of oscillatory cycles in the field before it hits the black hole singularity. The 
number of these observed cycles depends on the multipole index $\ell$ of the field and also whether 
the black hole is Schwarzschild or Kerr. We also found an indication of the formation of an 
asymptotic instability at the horizon of an extremal Kerr black hole. 
   
In future work, we would like to perform similar investigations of other matter fields, 
including the electromagnetic and gravitational cases. We also plan to explore the behavior 
of non-axisymmetric fields in a similar context.

{\em Acknowledgments}---We thank Lior Burko and Amos Ori for helpful discussions and for giving us 
feedback on a previous version of this paper. I.T. and G.K. acknowledges research support from NSF 
Grants No. PHY-1414440 and No. PHY–1606333, and from the U.S. Air Force agreement No. 10-RI-CRADA-09.

\medskip

\begin{thebibliography}{11}


\bibitem{GWPRL1}
B.~P.~Abbott {\it et al.} (LIGO and Virgo Scientific Collaborations)
Phys. Rev. Lett. {\bf116}, 061102 (2016).

\bibitem{GWPRL2}
B.~P.~Abbott {\it et al.} (LIGO and Virgo Scientific Collaborations)
Phys. Rev. Lett. {\bf116}, 241103 (2016).

\bibitem{price} R. H. Price, Phys. Rev. D {\bf 5}, 2419 (1972).

\bibitem{barack} L. Barack and A. Ori, Phys. Rev. Lett. {\bf 82}, 4388 (1999); L. Barack, Phys. Rev. D {\bf 61}, 024026 (2000).

\bibitem{hod} S. Hod, Phys. Rev. D {\bf 58}, 104022 (1998); S. Hod, Phys. Rev. D {\bf 60}, 104053 (1999); 
S. Hod, Phys. Rev. D {\bf 61}, 024033 (2000); S. Hod, Phys. Rev. Lett. {\bf 84}, 10 (2000); S. Hod, Phys. Rev. D {\bf 61}, 064018 (2000).

\bibitem{tails} L.M. Burko and G. Khanna, Class. Quant. Grav. {\bf 26}, 015014 (2009);
L. M. Burko and G. Khanna, Class. Quant. Grav. {\bf 28}, 025012 (2011);
I. Racz and G. Z. Toth, Class. Quant. Grav. {\bf 28}, 195003 (2011);
M. Jasiulek, Class. Quant. Grav. {\bf 29}, 015008 (2012);
T. Spilhaus and G. Khanna, preprint arXiv:1312.5210;
A.~Zengino\v{g}lu, G. Khanna and L. M. Burko, Gen. Rel. Grav. {\bf 46}, 1672 (2014);
L. M. Burko and G. Khanna, Phys. Rev. D {\bf 89}, 044037 (2014).

\bibitem{ori} E. Poisson and W. Israel, Phys. Rev. Lett {\bf 63}, 1663 (1989); A. Ori, Phys. Rev. Lett. {\bf 68}, 2117 (1992). 

\bibitem{brady} P. Brady and J. Smith, Phys. Rev. Lett {\bf 75}, 1256 (1995); L.M. Burko, Phys. Rev. Lett {\bf 79}, 4958 (1997).

\bibitem{cauchy} L. M. Burko, G. Khanna, A.~Zengino\v{g}lu, Phys. Rev. D {\bf 93}, 041501(R) (2016)

\bibitem{eilon} D. Marolf, A. Ori, Phys. Rev. {\bf D} 86, 124026 (2012); E. Eilon, A. Ori, Phys. Rev. {\bf D} 94, 104060 (2016).

\bibitem{others} M. Dafermos, Comm. Math. Phys. {\bf 332}, 729 (2014);
M. Dafermos, I. Rodnianski and Y. Shlapentokh-Rothman, Communications in Mathematical Physics Online first (2016);
A. Franzen, Communications in Mathematical Physics {\bf 343}, 601 (2014);
P. Hintz, preprint arXiv:1512.08003;
J. Luk and S.-J. Oh, preprint arXiv:1501.04598 (2015);
J. Luk and S.-J. Oh, Journal of Functional Analysis {\bf 271}(7), 1948 (2016); D. Philipp, V. Perlick, preprint arXiv:1503.08101.

\bibitem{ori92}  A. Ori, Phys. Rev. Lett. {\bf 83}, 5423 (1999). 

\bibitem{aretakis} S. Aretakis, Class. Quantum Grav. {\bf 30} 095010 (2013). 

\bibitem{gralla} M. Casals, S. E. Gralla, P. Zimmerman, Phys. Rev. D {\bf 94}, 064003 (2016).

\bibitem{teuk} S. Teukolsky, Astrophys. J. {\bf 185} 635 (1973).

\bibitem{ptc} M. Campanelli {\it et al}, Class. Quantum Grav. {\bf 18}, 1543 (2001).

\bibitem{anil} A. Zengino\v{g}lu, Class. Quant. Grav. {\bf 25}, 145002 (2008); 
A. Zengino\v{g}lu, Class. Quant. Grav. {\bf 25}, 175013 (2008); A. Zengino\v{g}lu, Class. Quant. Grav. {\bf 25}, 195025 (2008);
A. Zengino\v{g}lu, D. Nunez, S. Husa, Class. Quant. Grav. {\bf 26}, 035009 (2009); 
A. Zengino\v{g}lu, M. Tiglio, Phys. Rev. D {\bf 80}, 024044 (2009); A. Zengino\v{g}lu, Class.  Quant. Grav. {\bf 27}, 045015 (2010);
A. Zengino\v{g}lu,  J. Comput. Phys. {\bf 230}, 2286 (2011); A. Zengino\v{g}lu, Phys. Rev. D {\bf 83}, 127502 (2011).

\bibitem{hyper} A.~Zengino\v{g}lu and G.~Khanna, Phys.~Rev.~X {\bf 1}, 021017 (2011). Note that a 
different approach towards compactification was used therein -- a hyperboloidal compactified 
{\em layer} was attached to the outer part of a Boyer-Lindquist coordinates based computational 
grid.

\bibitem{harms} E.~Harms, S.~Bernuzzi, A.~Nagar, and A.~Zengino\v{g}lu, Class. Quantum Grav. {\bf 31}, 245004 (2014).

\bibitem{weno} Z. Grant, L. Isherwood, S. Gottlieb and G. Khanna, in progress. 

\bibitem{grg} A.~Zengino\v{g}lu, G. Khanna and L. M. Burko, Gen. Rel. Grav. {\bf 46}, 1672 (2014).

\bibitem{jsc} G. Khanna, J. Sci. Comput. {\bf 56}, 366 (2013). 

\bibitem{RHPthesis} R. H. Price, unpublished Ph.D. thesis, California Institute of Technology (1971).

\bibitem{ori98} A. Ori, Phys. Rev. D {\bf 57}, 2621 (1998).

\bibitem{BK} L. M. Burko, G. Khanna, Phys. Rev. D {\bf 89}, 044037 (2014).

\bibitem{ori2} A. Ori, Phys. Rev. D {\bf 58}, 084016 (1998).

\bibitem{next} L. Burko, G. Khanna, I. Thuestad, {\em in preparation}.  

\end{thebibliography}
\end{document}